\documentclass[preprint,aps,prl]{revtex4}
\usepackage{epsfig}
\usepackage{graphicx}
\input{epsf}

\begin{document}

\title{Entangled-Photon Imaging of a Pure Phase Object}

\author{Ayman F. Abouraddy,\footnote{Current address: M.I.T.
Research Laboratory of Electronics, Room 36-229, 77 Massachusetts
Avenue, Cambridge, Massachusetts 02139-4307; Electronic address:
raddy@mit.edu} Patrick R. Stone, Alexander V.
Sergienko,\footnote{Electronic address: alexserg@bu.edu} Bahaa E.
A. Saleh,\footnote{Electronic address: besaleh@bu.edu} and Malvin
C. Teich,\footnote{Electronic address: teich@bu.edu}}

\affiliation{Quantum Imaging Laboratory,\footnote{URL: $\:$
http://www.bu.edu/qil} Departments of Electrical \& Computer
Engineering and Physics, Boston University, Boston, Massachusetts
02215-2421, USA}

\date{\today}

\begin{abstract}
We demonstrate experimentally and theoretically that a coherent
image of a pure phase object may be obtained by use of a spatially
\textit{incoherent} illumination beam. This is accomplished by
employing a two-beam source of entangled photons generated by
spontaneous parametric down-conversion. Though each of the beams
is, in and of itself, spatially incoherent, the pair of beams
exhibits higher-order inter-beam coherence. One of the beams
probes the phase object while the other is scanned. The image is
recorded by measuring the photon coincidence rate using a
photon-counting detector in each beam. Using a reflection
configuration, we successfully imaged a phase object implemented
by a {\sc mems} micro-mirror array. The experimental results are
in accord with theoretical predictions.
\end{abstract}

\maketitle

\textit{Introduction}.--- It is well known in conventional optics
that a pure phase object cannot be imaged by illumination with a
spatially incoherent optical field \cite{Goodman68}.
``Coherence,'' in this case, refers to coherence in the
second-order \cite{MandelWolf95}. The question we ask is the
following: Can an optical field lacking second-order spatial
coherence, but endowed with higher-order coherence, be used to
measure the spatial distribution of a pure phase object? Clearly,
this is not possible by making use of conventional detectors,
which are sensitive to the optical intensity. Measurement of the
fourth-order coherence function has been used for imaging, as in
the Hanbury-Brown--Twiss interferometer \cite{MandelWolf95}, but
the imaging of a pure phase object can not be carried out using
such a configuration if a conventional thermal source of light is
used.

In this paper we demonstrate that the imaging of a phase object
can, in fact, be achieved by making use of a special light source
that is incoherent in second-order, but exhibits higher-order
coherence properties. Perhaps the most celebrated example of an
optical field exhibiting fourth-order coherence, in combination
with second-order \textit{in}coherence, is the twin-beam light
emitted by the process of spontaneous parametric down-conversion
(SPDC) in a nonlinear optical crystal \cite{Klyshko80}. Each of
the emitted beams lacks second-order spatial coherence (as well as
temporal coherence), but the two beams are endowed with inter-beam
higher-order spatial (and temporal) coherence, which is exhibited
via enhanced photon coincidences at directions for which photon
momentum is conserved
\cite{Ribeiro94_ApplOpt,Ribeiro94_PRA,Joobeur94_PRA,Joobeur96_PRA,Barbosa96_PRA,Saleh98_PRA}.
Such sources exhibit unique quantum-correlation features that have
generated considerable interest, particularly in the past few
years
\cite{Pittman95_PRA,Pittman96_PRA,Abouraddy01_PRL,Abouraddy02_JOSAB,Bennink02_PRL,Gatti03_PRL,Tan03_EurPhysJD,Howell03}.
Since photons are emitted in pairs, in an entangled quantum state,
imaging using such a light source has been referred to as
entangled-photon imaging. In this imaging technique, only one of
the beams interacts with the object; the photon coincidence rate
is measured much the same way as it is in the Hanbury-Brown--Twiss
interferometer.

Several experimental demonstrations and theoretical studies have
been directed at dissecting the capabilities and limitations of
entangled-photon imaging. On the theoretical side, it has been
shown that such a two-photon imaging system requires entanglement
in order to function as a coherent imaging system (i.e., as a
system capable of imaging a phase object) \cite{Abouraddy01_PRL}.
On the experimental side, the class of objects that has been
utilized in such demonstrations has been limited to that of
amplitude objects \cite{Pittman95_PRA, Pittman96_PRA}, e.g.,
objects with a spatially varying amplitude transmittance. Such
objects have usually taken the form of amplitude-modulating
transparencies inserted in a transmission configuration.

We proceed to provide an experimental demonstration of
entangled-photon imaging of a \textit{pure phase object} in a
\textit{reflection} configuration. The phase object is illuminated
by one of the (spatially incoherent) beams and the
photon-coincidence rate for detectors placed at both beams is
measured. This allows the formation of a coherent image of the
phase object. Phase objects are of special interest in quantum
information processing since they introduce a unitary operation
that is reversible (this is in contrast to amplitude masks of any
form).

\textit{Imaging configuration}.---The arrangement used in our
experiment is shown schematically in Fig. \ref{schematic}(a). The
source of light is spontaneous parametric down-conversion
generated in a nonlinear crystal ({\sc nlc}) pumped by a laser
beam ({\sc laser}). The downconverted photons are emitted in pairs
in two spatially separated beams (each of which is spatially
incoherent, as indicated above). One of the beams ({\sc probe})
impinges on the object, in our case a {\sc mems} micro-mirror
array configured so as to modulate the phase of the impinging
wavefront. Upon reflection from the object, the probe beam is
detected by a single detector {\sc d}$_{1}$ fitted with a fixed
small pinhole {\sc p}$_{1}$. The other beam ({\sc reference}) does
not interact with the object, and is simply directed to a  small
pinhole {\sc p}$_{2}$ and a detector {\sc d}$_{2}$. The terms
``{\sc probe}'' and ``{\sc reference}'' are used since they are
more descriptive in this case than the usual terminology: signal
and idler.  The coincidence rate of the photons detected by the
two detectors is measured  as detector {\sc d}$_{2}$ (together
with pinhole {\sc p}$_{2}$) is scanned. As will be shown
theoretically, and confirmed experimentally, if the intervening
optical system is appropriately designed, a coherent image of the
phase object may be obtained.

The system is configured such that the probe beam cannot, by
itself, generate an image of the phase object. First, detector
{\sc d}$_{1}$  is a single fixed detector lacking spatial
resolution. Second, the probe beam lacks second-order spatial
coherence. Third,  the distance $d_{b}$ between the object and the
detector {\sc d}$_{1}$ is deliberately chosen such that {\sc
d}$_{1}$ does not lie in the far-field, thus precluding the
formation of a far-field diffraction pattern of the phase object,
which could provide information about the phase spatial
distribution were the beam coherent.

The pure phase object we consider is characterized by a spatially
varying unimodular complex amplitude transmittance or reflectance
of the form $\exp[i\theta(\textbf{r})]$, where
$\theta(\textbf{r})$ is a real function of the position
$\textbf{r}$ in the object plane. Such an object could not be
imaged directly by using a conventional single-lens imaging system
satisfying the imaging condition and a conventional
intensity-sensitive detector.  Since such a system ideally
provides a geometric mapping between each point in the object
plane and a corresponding point in the image plane,  it yields no
phase information.  In the context of entangled-photon imaging,
such a system was implemented in Ref. \cite{Pittman95_PRA} for
imaging the intensity transmittance of an object.

In conventional coherent optics, a phase object is typically
imaged either interferometrically or by use of an optical spatial
Fourier-transform system, which converts the phase distribution of
the optical wave front at the object plane into a spatially
varying amplitude distribution that is detectable by an
intensity-sensitive detector.  Two examples of Fourier-transform
systems are (1) a single lens 2--$f$ system, and (2) free-space
propagation in the Fraunhofer regime, commonly known as the
far-field \cite{Saleh_book}. In our entangled-photon imaging
system, we have implemented a Fourier-transform configuration
based on lensless propagation to the far-field, a system similar
to that used in Ref. \cite{Strekalov95_PRL}. The exact Fourier
transform is achieved only in the Fraunhofer-diffraction region,
which necessitates traveling a prohibitively long distance
\cite{Saleh_book}. The distances used in our experiment actually
place the image in the Fresnel-diffraction region.

\textit{Theory}.---A theoretical expression for the photon
coincidence rate in the above imaging configuration is obtained by
using the formalism developed in Ref. \cite{Saleh00_PRA}. The
coincidence rate of photon pairs at points $\textbf{x}_{1}$ and
$\textbf{x}_{2}$, in the planes of detectors {\sc d}$_{1}$ and
{\sc d}$_{2}$, is proportional to the fourth-order coherence
function of the optical fields
\begin{equation}
G^{(2)}(\textbf{x}_{1},\textbf{x}_{2})=\left|\int\int
d\textbf{x}\,d\textbf{x}'\,\phi(\textbf{x},\textbf{x}')
\,h_{1}(\textbf{x}_{1},\textbf{x})\,h_{2}(\textbf{x}_{2},\textbf{x}')\right|^{2}.
\label{eq1}
\end{equation}
Here $h_{1}$ and $h_{2}$ are the impulse response functions
describing the optical systems that the probe and reference
photons traverse from the crystal to {\sc d}$_{1}$ and {\sc
d}$_{2}$, respectively. In the arrangement outlined here, both
$h_{1}$ and $h_{2}$ represent free-space propagation, and $h_{1}$
includes the phase object. The quantum state of light produced in
the process of SPDC is represented by the state vector
\begin{equation}
|\Psi\rangle=\int\int
d\textbf{x}\,d\textbf{x}'\,\phi(\textbf{x},\textbf{x}')\,|1_{\textbf{x}},1_{\textbf{x}'}\rangle;
\end{equation}
the state function $\phi$ is normalized so that $\int\int
d\textbf{x}\,d\textbf{x}'\,|\phi(\textbf{x},\textbf{x}')|^{2}=1$.
The state function is related to the physical parameters of the
SPDC source through the relation
\begin{equation}
\phi(\textbf{x},\textbf{x}')=\int d\textbf{y}\,
E_{p}(\textbf{y})\,\xi(\textbf{x}-\textbf{y},\textbf{x}'-\textbf{y}),
\end{equation}
where $E_{p}(\textbf{x})$ is the pump field spatial profile,;
$\xi(\textbf{x},\textbf{x}')$ is the Fourier transform of the
phase-matching function
$\tilde{\xi}(\textbf{\textbf{q}}_{1},\textbf{\textbf{q}}_{2})=
\textrm{sinc}(\frac{l}{2\pi}\Delta)\,\exp(-i\frac{l}{2}\Delta)$;
$l$ is the thickness of the nonlinear crystal; $\textbf{q}_{1}$
and $\textbf{q}_{2}$ are the transverse momenta of the probe and
reference beams; and $\Delta(\textbf{q}_{1},\textbf{q}_{2})$ is
the mismatch in longitudinal momenta of the pump, probe, and
reference beams. These formulas all assume that only a narrow band
of wavelengths, centered around the degenerate wavelength, is
allowed, as is usually imposed by the use of interference filters
in the optical arrangement.

Coherent image formation in this configuration may be understood
by noting that the left hand side of Eq.(\ref{eq1}) is
mathematically identical to the coherent image at $x_{2}$ of a
point source at $x_{1}$ illuminating an optical system composed of
a cascade of systems with impulse response functions
$h_{1}(\textbf{x},\textbf{x}_{1})$,
$\phi(\textbf{x}',\textbf{x})$, and
$h_{2}(\textbf{x}_{2},\textbf{x}')$.   This is physically
identical to a point source located at the point detector {\sc
d}$_1$ emitting light traveling backward through $h_{1}$ towards
the nonlinear crystal, which serves as a special mirror reflecting
the wave through $h_{2}$, and creating a coherent image in the
plane of detector {\sc d}$_2$. This advanced-wave interpretation,
pioneered by D.~N. Klyshko \cite{klyshko88}, provides a rationale
for the formation of a Fourier-transform image of the phase object
in the far field.

The effect of the finite size of {\sc p}$_{1}$ may be accommodated
by integrating the variable $\textbf{x}_{1}$ over the area of {\sc
p}$_{1}$,
\begin{equation}\label{coincidence}
C(\textbf{x}_{2})=\int_{\textrm{p}_{1}}
d\textbf{x}_{1}\,G^{(2)}(\textbf{x}_{1},\textbf{x}_{2}).
\end{equation}
The quantity $C(\textbf{x}_{2})$ is the coincidence rate measured
when {\sc p}$_{2}$ is scanned, while {\sc p}$_{1}$ is held fixed,
and corresponds to the data collected in the experiment.  In view
of the advanced-wave interpretation, the finite size of {\sc
p}$_{1}$ introduces partial coherence into the imaging system and
may render the system effectively incoherent  \cite{Saleh00_PRA}.
The size of the pinhole {\sc p}$_{2}$, on the other hand, sets a
limit on the resolution of the scanned coherent image.

Calculating $C(\textbf{x}_{2})$ is difficult when all the physical
parameters of the configuration are considered. To simplify the
calculation we assumed that the nonlinear crystal is thin
($l\rightarrow 0$) and that the pump is a plane wave incident
normally on the crystal. This zeroth-order approximation
underestimates the width of the produced far-field image, since an
infinite spatial spectrum is implicit in the thin-crystal
approximation. We have therefore approximately accommodated the
finite width of the pump and finite crystal thickness by
multiplying the calculated $C(x_{2})$ by the measured conventional
far-field image of the reference obtained from single-photon rates
at {\sc d}$_{2}$, which is independent of the object and  depends
only on the parameters of the pump beam and the nonlinear crystal.

\textit{Experiment}.---The pure phase object used in our
experiments comprised a 12$\times$12 array of gold-plated
micro-mirrors, each of dimension
300$\times300\,\mu\textrm{m}^{2}$. The height of each mirror, with
respect to a fixed datum, is altered via an electrostatic
potential. If one micro-mirror is pulled down from the datum a
distance $d$, the portion of the wavefront impinging on this
region  accumulates a phase $2\pi({2d}/{\lambda})$ relative to the
datum (which is taken to be phase $0$), after reflection from the
micro-mirror and double traversal of the distance $d$. For
example, light at $\lambda=812$ nm, such as that in our
experiment, when reflected from a micro-mirror pulled back a
distance $d=200$ nm, accumulates a phase of approximately $\pi$
radians.

We conducted experiments using three distinct phase distributions:
(a) zero phase everywhere (flat mirrors); (b) a single line of
micro-mirrors pulled down to implement a phase of $\pi$; and (c)
two lines of micro-mirrors pulled down to implement a phase of
$\pi$ separated by an undisturbed line of phase zero. All three
distributions are independent of one dimension of the array, and
are thus effectively one-dimensional distributions ($\textbf{r}
\to x$). This is helpful since it enables us to integrate along
the uniform direction at {\sc d}$_{1}$ and {\sc d}$_{2}$.

A more detailed view of the experimental arrangement is
illustrated in Fig. \ref{experiment}. The pump was the 406-nm line
of a cw Kr-ion laser with a power of 30 mW. The nonlinear crystal
was a 1.5-mm-thick BBO crystal cut for collinear (probe and
reference emitted into the same beam), degenerate, type-I (probe-
and reference have the same polarization) SPDC. We chose a type-I
collinear configuration, rather than a type-II (probe- and
reference with orthogonal polarizations) collinear configuration
such as that used in Refs.~\cite{Pittman95_PRA} and
\cite{Pittman96_PRA}, or the type-I non-collinear configuration
used in Ref.~\cite{Ribeiro94_PRA}. The advantage of the type-I
collinear configuration is that the two down-converted beams are
emitted in the same circularly symmetric spatial mode. This is
useful for carrying out imaging experiments since artifacts
arising from differences between the spatial distributions of the
two beams, as well as peculiarities of beam shape, are eliminated.
However, half of the photon-pair flux is lost in this
configuration, as will be elaborated upon below.

The pump (extraordinary polarization) was separated from the
down-converted photons (ordinary polarization) by means of a pair
of Glan-Laser polarizing beam splitters ({\sc gl}$_{1}$ and {\sc
gl}$_{2}$), placed before and after the nonlinear crystal,
respectively. A long-pass colored glass filter ({\sc cgf}) of
cutoff wavelength 560 nm was used to further separate away the
pump. The photon pairs then permitted to impinge on a
nonpolarizing sheet beam splitter ({\sc bs}). As a result, half of
the pairs are separated into the two output beams whereas the
other half of the pairs emerge together at the same output port
and thus fail to contribute to coincidences; this accounts for the
50\% reduction of photon flux mentioned above.

The beam reflected from the beam splitter impinges on the phase
object and from there is reflected to a single-photon-detector
module (fiber-coupled EG$\&$G SPCM-AQR-15). The detector is
preceded by a vertical slit {\sc p}$_1$ of width 1.4 mm and an
interference filter {\sc f}$_1$ (centered at 800 nm with a
bandwidth of 66 nm) to eliminate any remaining pump photons. The
detected photons are integrated along the direction parallel to
the slit. The beam transmitted through the beam splitter, which
does not interact with the object, is directed to an identical
detection unit (vertical slit {\sc p}$_2$ and filter {\sc f}$_2$).
However this detector is mounted on a computer-controlled stage
that permits scanning of the beam. The electrical pulses from the
two detectors are sent to a coincidence circuit ($\otimes$) with a
2-nsec timing window.

The distance from the {\sc nlc} through the {\sc bs} to the
micro-mirrors is $d_a=1.17$ m; the distance from the micro-mirrors
to the detector {\sc d}$_1$ is $d_b=1.98$ m; the distance from the
{\sc nlc}, reflecting from the {\sc bs}, to {\sc d}$_2$ is
$d_2=3.96$ m. For this configuration, the diffraction pattern is
formed at a distance $d_2+d_a$ from the mirrors: from the mirrors
to the {\sc bs} to the {\sc nlc}, back to the {\sc bs} and then on
to {\sc d}$_2$.

 \textit{Results}.---The results are displayed in Fig.~\ref{data}
for the three phase objects on which we conducted imaging
experiments. In each case, the coincidence counts measured by
detectors {\sc d}$_{1}$ and {\sc d}$_{2}$ is plotted as a function
of the scanned position $x_2$ of detector {\sc d}$_{2}$. The
experimental results (open circles) are seen to match the
theoretical predictions based on the approximate model (solid
curves) reasonably well, although the theoretical profiles are
slightly wider.  We have also recorded the single-photon counts
collected from {\sc d}$_{2}$, independently of the counts recorded
from {\sc d}$_{1}$. This measure is, of course, completely
independent of the object; it represents the far-field pattern of
the reference beam and thus depends only on the physical
parameters of the pump and nonlinear crystal. The first object has
a uniform phase distribution, $\theta(x)=0$, which represents a
highly reflecting, uniform mirror of finite aperture. The
coincidence profile shown in Fig.~\ref{data}(a) is simply the
Fourier transform of this object, which is the diffraction pattern
of the object aperture.

The second object has a phase distribution in the form of a single
strip of phase $\pi$ in a uniform background of phase zero.  We
call this object  a \textit{phase slit}. The measured coincidence
image presented in Fig.~\ref{data}(b) has a double-peaked profile
that is the Fresnel transform (approximately the Fourier
transform) of the object phase distribution.   This profile is
dramatically different from that associated with the usual
\textit{amplitude slit}, which has a single central peak and
smaller side lobes.

The third object is a double phase slit. The measured coincidence
profile displayed in Fig.~\ref{data}(c)  is qualitatively similar
to that for the single phase slit illustrated in
Fig.~\ref{data}(b), but has  a wider and deeper dip, and also a
lower peak value. There are two reasons for this: (1) The
diffraction pattern for the double phase slit is more spread out,
and since no photons are absorbed by the phase object, the height
of the distribution must be lower; (2) The rate of single photons
detected by {\sc d}$_{1}$ is smaller. The pattern reflected
towards {\sc d}$_{1}$ is wider, and since the pinhole has the same
width, fewer photons are collected for the double phase slit,
thereby resulting in a further reduction of coincidence counts for
this case. We have recorded the single-photon rate at both
detectors in the two cases. It is gratifying that the
incorporation of this additional reduction factor into the
theoretical calculation leads to a ratio of heights of the
theoretical patterns that match experiment.

The diffraction profile of a double-phase-slit object is
dramatically different from that of a conventional double-slit
amplitude object; the latter exhibits a single central peak with
smaller side lobes, rather than a double-peaked distribution. For
the same configuration and dimensions, the amplitude object
produces a substantially wider diffraction pattern. Note, however,
that a ``double-strip'' amplitude object (an aperture that
transmits light everywhere except at two parallel strips) has a
diffraction pattern similar to that of a double phase slit of the
same dimensions, but it has far less visibility (it is well known
in phase lithography that the maximum contrast produced by slit
modulation is obtained for $\pi$-phase-shift slits).

\textit{Conclusion}.---We have experimentally demonstrated that a
coherent image of a reflective pure phase object may be obtained
by using a spatially incoherent probe beam.  This is accomplished
by use of an auxiliary  reference beam  that does not interact
with the object. Generated by spontaneous parametric
down-conversion, each of the two beams is spatially incoherent but
together they exhibit inter-beam higher-order coherence, by virtue
of the fact that the source emits photon pairs in an entangled
state. By measuring the photon coincidence rate, using a
single-photon detector in each of the two beams, we observe the
fourth-order cross-coherence function for the reflected probe beam
and the reference beam. This contains an image of the phase object
identical to that measurable with coherent light. The experiments
we carried out include measurements of the Fresnel (approximately
Fourier) transform of simple phase objects, but other coherent
images may be similarly measured. This includes phase-contrast
imaging, which may be realized by splitting the probe beam into
two laterally displaced probe beams transmitted through the
object.  It also includes holography, which may be realized by
splitting the probe beam into a reference and an object beam as in
conventional holography \cite{Abouraddy01_OptExp}.

\textit{Acknowledgments}.---This work was supported by the
National Science Foundation; by the Center for Subsurface Sensing
and Imaging Systems (CenSSIS), an NSF Engineering Research Center;
and by the David \& Lucile Packard Foundation. We thank N.
Vamivakas and J. Hofman for technical assistance. We are grateful
to the Boston Micromachines Corporation and T. G. Bifano for
providing us with the {\sc mems} micro-mirror array.

\newpage

\begin{figure}[htbp]
\caption{(a) Schematic of the experimental arrangement for the
quantum imaging of a reflective pure phase object ({\sc mems}).
{\sc nlc} represents a nonlinear crystal, {\sc p}$_{1}$ and {\sc
p}$_{2}$ are pinholes, {\sc d}$_{1}$ and {\sc d}$_{2}$ are
detectors, $\otimes$ represents an electronic coincidence circuit,
and the $d_{[\cdot]}$ are distances.} \label{schematic}
\end{figure}

\begin{figure}[htbp]
\caption{(a) Actual experimental arrangement for the quantum
imaging of a reflective pure phase object ({\sc mems}). {\sc nlc}
represents the nonlinear crystal, {\sc cgf} stands for a colored
glass filter, {\sc gl}$_{1}$ and {\sc gl}$_{2}$ are two
orthogonally oriented Glan-Laser polarizing beam splitters, {\sc
bs} is a nonpolarizing beam splitter, {\sc p}$_{1}$ and {\sc
p}$_{2}$ are vertical slits, {\sc f}$_{1}$ and {\sc f}$_{2}$ are
interference filters, {\sc d}$_{1}$ and {\sc d}$_{2}$ are
single-photon detectors, and $\otimes$ represents an electronic
coincidence circuit.} \label{experiment}
\end{figure}

\begin{figure}[htbp]
\begin{center}
\caption{Experimental coincidence counts (open circles) measured
from {\sc d}$_{1}$ and {\sc d}$_{2}$ as the position $x_2$ of
detector {\sc d}$_{2}$ is scanned. The collection time is 80 sec
per point. The solid curves are the theoretical predictions. (a)
Object with uniform phase zero; (b) Single-slit object with phase
$\pi$; (c) Double-slit object with phase $\pi$ separated by a line
of phase $0$.} \label{data}
\end{center}
\end{figure}

\end{document}